\documentstyle[prb,aps]{revtex}

\def\beq{\begin{eqnarray}}
\def\eeq{\end{eqnarray}}

\begin{document}

\title{Thermoelectric Effect in High-$T_c$ Superconductors: the Role of
Density of States Fluctuations}
\author{D. V. Livanov}
\address{ Moscow Institute for Steel and Alloys, \\
Leninsky prospect 4, 117936 Moscow, Russia.\\}
\author{A. A. Varlamov}
\address{Department of Physics "A. Volta", University of Pavia\\
Strada Nuova 65, 27100 Pavia, Italy\\}
\author{F. Federici}
\address{Department of Physics, University of Florence\\
Largo E. Fermi 2, 50125 Florence, Italy.}
\date{today}
\maketitle

1. The problem of thermoelectric effect in fluctuation regime
has been attracting the attention of theoreticians during more than
twenty years, since the paper of Maki \cite{Maki74}.
The main question which should be answered is whether the correction to
thermoelectric coefficient $\beta$ has the same temperature singularity in 
vicinity of critical temperature $T_{\rm c}$ as the correction to electrical 
conducticity $\sigma$ or not.
In the paper of Maki \cite{Maki74} the only logarithmically divergent
contribution was predicted in two-dimentional (2D) case and its
sign was found to be opposite to the sign of the normal state
thermoelectric coefficient $\beta_0$. Later on, in a number of papers
\cite{VL90,R91,K93} it was claimed that temperature singularity
of fluctuation correction to $\beta$ is the same as it is for $\sigma$
($\propto(T-T_{\rm c})^{-1}$ in 2D). Finally,
Reizer and Sergeev \cite{RS94} have recently revised the problem using both
quantum kinetic equation and linear response methods and have
shown that,
in the case of isotropic electron spectrum,
strongly divergent contributions \cite{VL90,R91,K93} are cancelled
out for any dimensionality, while the final result has the same
logarithmic singularity as it was found by Maki, but the opposite sign.
We should emphasize that in all papers cited above the only Aslamazov-Larkin
(AL) contribution was taken into account, while the anomalous
Maki-Thompson (MT) term was shown to be absent \cite{VL90,RS94}.
It was mentioned \cite{RS94} that the non-correct evaluation of interaction 
corrections to heat-current operator in Refs. \cite{VL90,R91,K93}
produced the erroneously large
terms which really are cancelled out within the adequate procedure.
Due to this strong cancellation the AL term turns out to be less
singular compared with corresponding correction to conductivity \cite{RS94}.

From the other side, now it is well established that in every case where
the senior AL and MT fluctuation corrections are suppressed by some
reason, the contribution connected with fluctuation renormalization
of one-electron density of states (DOS) can become important.
As examples we can mention
$c$-axis fluctuative transport \cite{ILVY93,BDKLV93}, NMR relaxation rate
\cite{RV94} and infrared optical conductivity \cite{fv96}.
In this communication we show that the analogous situation
takes place also in the case of thermoelectric coefficient.
In what follows we study the DOS contribution to the thermoelectric
coefficient of superconductors with an arbitrary impurity concentration above
$T_{\rm c}$.
We will be mostly interested in 2D case, but the generalization
to the case of layered superconductor will be done at the end.
We show that, although DOS term has the same temperature dependence as
AL contribution \cite{RS94}, it turns out to be the
leading fluctuation contribution both in clean and dirty cases
due to its specific dependence on electron mean free path.

2. We use units with $\hbar=c=k_{\rm B}=1$. We introduce the thermoelectric
coefficient $\beta$ in the framework of linear response theory as:
\beq
\beta=\lim_{\omega\rightarrow 0}\frac{{\rm Im}[Q^{\rm (eh)R}
(\omega)]}{T\omega}
\eeq
where $Q^{\rm (eh)R}(\omega)$ is the retarded Fourier component of the
correlation function of electric and heat current operators.
This correlation function in diagrammatic technique is represented by the two
exact electron Green's functions loop with two external field vertices,
the first, $-e{\bf v}$, associated with the electric current operator and the
second one, $\frac{i}2(\epsilon_n+\epsilon_{n+\nu}){\bf v}$, associated with
the heat current operator ($\epsilon_n=\pi T(2n+1)$ is fermionic Matsubara 
frequency and ${\bf v}=\partial\xi({\bf p})/\partial {\bf p}$ with $\xi$
being the quasiparticle energy).
Taking into account the first order of perturbation theory in Cooper
interaction and averaging over impurity configuration
one can find ten diagrams presented in Fig. 1.
The solid lines represent
$G({\bf p},\epsilon_n)=1/(i\tilde\epsilon_n-\xi({\bf p}))$,
the single-quasiparticle normal state Green's function averaged
over impurities which contains the scattering lifetime $\tau$
($\tilde\epsilon_n=\epsilon_n+1/2\tau {\rm sign}\epsilon_n$).
The shaded objects are the vertex impurity renormalization
$\lambda({\bf q}=0,\epsilon_n,\epsilon_{n^{\prime}})$ (see \cite{BDKLV93}).
The wavy line represents the fluctuation propagator
$L({\bf q},\Omega_k)$:
\beq
\label{prop}
L^{-1}({\bf q},\Omega_k)=-\rho\left[\ln \frac{T}{T_{\rm c}}+\psi\left(
\frac12+\frac{|\Omega_k|}{4\pi T}+\frac{2\eta q^2}{\pi^2}
\right)-\psi\left(\frac12\right)\right]
\eeq
where
$$
\eta =-\frac{v_{\rm F}^2\tau^2}{2}\left[\psi\left(\frac12+\frac{1}{4\pi
T\tau}\right)-\psi\left(\frac12 \right)-\frac{1}{4\pi T\tau}\psi^\prime
\left(\frac12\right) \right]
$$
is a positive constant which enters into the current expression in
Ginzburg-Landau theory in 2D case ($\rho$ is one-electron density of
states and $\psi(x)$ and $\psi^\prime(x)$ are
digamma function and its derivative, respectively).
The first diagram
describes the AL contribution to thermoelectric coefficient
and was calculated in \cite{RS94} with electron-hole asymmetry factor
taken into account in fluctuation propagator.
Diagrams 2-4 represent Maki-Thompson contribution.
As it was mentioned in Refs. \cite{VL90,RS94}, neither anomalous nor regular
parts of this diagram contribute to $\beta$ in any order of electron-hole
asymmetry.
In what follows we will discuss the contribution from diagrams 5-10 which
describes the correction to $\beta$ due to DOS renormalization.

For diagrams 5 and 6 we have
\beq
\label{Q12}
Q^{(5+6)}(\omega_\nu)&=&-2eT\sum_{\Omega_k}\int (d{\bf q})
L({\bf q},\Omega_k)T\sum_{\epsilon_n}\frac{i\left(
\epsilon_{n+\nu}+\epsilon_n\right)}{2}\int (d{\bf p})
v^2\times\nonumber\\
\nonumber\\
&\times&\left[\lambda^2(\epsilon_n,-\epsilon_n)
G^2\left({\bf p},\epsilon_n\right)G\left({\bf q}-{\bf p},-\epsilon_n\right)
G\left({\bf p},\epsilon_{n+\nu}\right)+\right. \\
\nonumber\\
&+&\left.\lambda^2(\epsilon_{n+\nu},-\epsilon_{n+\nu})
G^2\left({\bf p},\epsilon_{n+\nu}\right)
G\left({\bf q}-{\bf p},-\epsilon_{n+\nu}\right)
G\left({\bf p},\epsilon_{n}\right)\right].\nonumber
\eeq
(We use the shorthand notation $(d{\bf q})= d^dq/(2\pi)^d$, where
$d$ is dimentionality).
Evaluating Eq. (\ref{Q12}) one naturally obtains zero result without
taking into account the electron-hole asymmetry. The first possible
source of this factor is contained in fluctuation propagator and
was used in \cite{RS94} for AL diagram. Our calculations
show that for DOS contribution this correction to fluctuation
propagator results in non-singular correction to $\beta$ in 2D case and
can be neglected. Another source of electron-hole asymmetry is connected with
expansion of energy-dependent functions in power of $\xi/E_{\rm F}$
near Fermi level performing ${\bf p}$-integration in Eq. (\ref{Q12})
($E_{\rm F}$ is the Fermi energy):
\beq \label{exp}
\rho v^2(\xi)=\rho v^2(0)+
\xi\left[\frac{\partial(\rho v^2(\xi))}{\partial\xi}
\right]_{\xi=0}.
\eeq
Only second term in Eq. (\ref{exp}) contributes to thermoelectric
coefficient.
Contribution of diagrams 7 and 8 can be calculated in analogous way.
Diagrams 9-10 do not give any singular contribution to thermoelectric
coefficient due to the vector character of external vertices and as a
result an additional $q^2$ factor appears after ${\bf p}$-integration.
The same conclusion concerns MT-like diagram 3-4.

Performing integration over $\xi$ we find the contribution of the
important diagrams 5-8 in the form
\beq \label{Q12_c}
Q^{(5-8)}(\omega_\nu)&=&
\frac{eT^2}{4}\left[\frac{\partial(\rho v^2(\xi))}{\partial\xi}
\right]_{\xi=0}\int (d{\bf q}) L({\bf q},0)(\Sigma_1+
\Sigma_2 + \Sigma_3),
\eeq
where we have separated sums over
semi-infinite ($]-\infty, -\nu -1]$, $[0, \infty[$) and
finite ($[-\nu, -1]$) intervals :
\beq\label{sums}
\Sigma_1&=&2\sum_{n=0}^{\infty}\frac{2\epsilon_n+\omega_{\nu}}{2\tilde
\epsilon_n+\omega_{\nu}}\left(\frac{\tilde\epsilon_n+\omega_{\nu}}
{(\epsilon_n+\omega_{\nu})^2}+
\frac{\tilde\epsilon}{\epsilon_n^2}\right),\nonumber\\
\nonumber \\
\Sigma_2&=& \frac1{(1/\tau+\omega_{\nu})^2}\sum_{n=-\nu}^{-1}
(2\epsilon_n+\omega_{\nu})^2\left(\frac{\tilde\epsilon_{n+\nu}}
{\epsilon_{n+\nu}^2}-\frac{\tilde\epsilon_n}{\epsilon_n^2}\right)\\
\nonumber\\
\Sigma_3&=&
(1+\omega_{\nu}\tau)\sum_{n=-\nu}^{-1}(2\epsilon_n+\omega_{\nu})
\left(\frac1
{\epsilon_{n+\nu}^2}-\frac1{\epsilon_n^2}\right).
\nonumber
\eeq
$\Sigma_1$ and $\Sigma_2$ are associated with diagram 5-6, while
$\Sigma_3$ with diagram 7-8.
Calculating sums (\ref{sums}) we are interested in terms which are linear
in external frequency $\omega_{\nu}$. Sum $\Sigma_1$ turns out to be an
analytical function of $\omega_{\nu}$ and it is enough to expand it
in Taylor series after analytical continuation $\omega_{\nu}\rightarrow
-i \omega$. The last two sums over finite intervals require more
attention because of their nontrivial $\omega_{\nu}$-dependence and before
analytical continuation they have to be calculated rigorously.
As a result:
\beq
\Sigma^R_1=\frac{i\omega}{4T^2} \ ; \
\Sigma^R_2=-\frac{2i\omega\tau}{\pi T}\ ;\
\Sigma^R_3=-\frac{i\omega}{2T^2}
\eeq

Finally, we perform integration over
${\bf q}$ and the total contribution associated with DOS
renormalization in 2D case takes the form:
\beq\label{final1}
\beta^{\rm DOS}=\frac1{8\pi^2}\frac{eT_c}{v_{\rm F}^2\rho}
\left[\frac{\partial(v^2\rho)}{\partial\xi}\right]_{\xi=0}
\ln\left(\frac{T_{\rm c}}{T-T_{\rm c}}\right) \kappa(T_c\tau),
\eeq
\beq
\label{kappa}
\kappa(T\tau)&=&
-\displaystyle{\frac{1+\displaystyle{\frac{\pi}{8T\tau}}}
{T\tau\left[\psi\left(\displaystyle{\frac12}+\displaystyle{\frac{1}{4\pi
T\tau}}\right)-\psi\left(\displaystyle{\frac12} \right)-
\displaystyle{\frac{1}{4\pi T\tau}}\psi^\prime
\left(\displaystyle{\frac12}\right)\right]}}\nonumber\\
\\
&=&\cases{\displaystyle{\frac{8\pi^2}{7\zeta(3)}}
T\tau \approx 9.4 T\tau\ \ \ & for $T\tau \gg 1$ \cr
\displaystyle{\frac1{T\tau}}\ \ \ & for $T\tau \ll 1$}
\nonumber
\eeq
To generalize this result to the important case of
layered superconductor one has to replace
$\ln(1/\epsilon)\rightarrow\ln[2/(\sqrt{\epsilon}+\sqrt{\epsilon+r})]$
($\epsilon=(T-T_{\rm c})/T_{\rm c}$ and $r$ is an anisotropy parameter
\cite{BDKLV93})
and to multiply Eq. (\ref{final1}) by $1/p_{\rm F} s$, where
$s$ is the interlayer distance. In the limiting case
of 3D superconductor ($r\gg\epsilon$) both AL \cite{RS94} and
DOS contributions are not singular.

3. Comparing Eq. ({\ref{final1})
with the results of \cite{RS94}
for AL contribution, we conclude, that in both limiting cases of clean
and dirty systems the decrease of $\beta$ due to fluctuation DOS
renormalization dominates the thermoelectric transport due to
AL process. Really, the total relative correction to thermoelectric
coefficient in the case of 2D superconducting film of thickness $s$
can be written in the form:
\beq
\frac{\beta^{\rm DOS}+\beta^{\rm AL}}{\beta_0}
=-0.17 \frac{1}{E_{\rm F}\tau}\frac{1}{p_{\rm F} s}\ln\left(\frac{T_{\rm c}}
{T-T_{\rm c}}\right)
\left[0.5\kappa(T_{\rm c}\tau)+5.3\ln\frac{\Theta_{\rm D}}{T_{\rm c}}\right],
\eeq
where the first term in square brackets corresponds to the DOS
contribution (\ref{final1}) and the second term describes the AL contribution
from Ref. \cite{RS94} ($\Theta_{\rm D}$ is Debye temperature).
Assuming $\ln(\Theta_D/T_{\rm c})\approx 2$ one finds that DOS
contribution dominates AL one for any value of impurity concentration:
$\kappa$ as a function of $T\tau$ has a minimum at
$T\tau \approx 0.3$ and even at this point DOS term is twice larger.
In both limiting cases $T\tau \ll 1$ and $T\tau \gg 1$ this
difference strongly increases.

The temperature and impurity concentration dependencies of
fluctuation corrections to $\beta$ can
be evaluated through a simple qualitative consideration.
The thermoelectric coefficient may be estimated through the electrical
conductivity $\sigma$ as
$\eta\sim(\epsilon^\ast/eT)f_{\rm as}\sigma$,
where $\epsilon^\ast$ is the characteristic energy involved in
thermoelectric transport and $f_{\rm as}$ is the electron-hole asymmetry
factor, which is defined as the ratio of the difference between numbers of
electrons and holes to the total number of particles.
Conductivity can be estimated as $\sigma \sim e^2 {\cal
N}\tau^{\ast}/m$, where ${\cal N}$, $\tau^{\ast}$ and $m$ are the
density, lifetime and mass of charge (and heat) carriers, respectively.
In the case of AL contribution the heat carriers are
nonequlibrium Cooper pairs with energy $\epsilon^\ast \sim T-T_{\rm c}$
and density ${\cal N}\sim p_{\rm F}^d \frac{T}{E_{\rm F}}\ln\frac{T_{\rm c}}
{T-T_{\rm c}}$ and
characteristic time, given by Ginzburg-Landau time
$\tau^\ast \sim \tau_{GL} = \frac{\pi}{8(T-T_{\rm c})}$.
Thus in 2D case $\Delta\eta^{\rm AL}\sim(T-T_{\rm c})/(eT_{\rm c})
f_{\rm as}\Delta\sigma^{\rm AL} \sim e f_{\rm as} \ln\frac{T_{\rm c}}
{T-T_{\rm c}}$.
One can easily get that the
fluctuation correction due to AL process is less singular (logarithmic
in 2D case) with respect to the corresponding correction to conductivity
and does not depend on impurity scattering \cite{RS94}.

The analogous consideration of the single-particle DOS contribution
($\epsilon^\ast \sim T$, $\tau^\ast \sim \tau$) evidently
results in the estimate
$\beta\sim e f_{\rm as}T_{\rm c}\tau\ln{\frac{T_{\rm c}}{T-T_{\rm c}}}$
which  coincides with (\ref{final1}) in clean case.
The dirty case is more sofisticated because the fluctuation
density of states renormalization strongly depends on the character of the
electron motion, especially in the case of diffusive
motion \cite{CCRV90}. The same density of states redistribution in
the vicinity of Fermi level directly enters into the rigorous expression
for $\beta$ and it is not enough to write the fluctuation Cooper
pair density $\cal N$ but it is necessary to take into account
some convolution with $\delta\rho_{\rm fl}(\epsilon)$.
This is what was actually done in the previous calculations.

Experimentally, although Seebeck coefficient $S=-\eta/\sigma$ is probably the
easiest to measure among thermal transport coefficients, the comparison
between experiment and theory is complicated by the fact that $S$ cannot
be calculated directly; it is rather a composite quantity of electrical
conductivity and thermoelectric coefficient. As both $\eta$ and $\sigma$
have corrections due to superconducting fluctuations, total correction
to Seebeck coefficient is given by
\begin{equation} \label{S}
\Delta S=S_0\left( \frac{\Delta \beta}{\beta_0}-\frac{\Delta \sigma}{\sigma_0}
\right)
\end{equation}
Both these contributions provide a positive correction $\Delta\beta$, thus
resulting in the decrease of the absolute value of $S$ at the edge of
superconducting transition ($\Delta \beta/\beta_0 < 0$).
As for fluctuation correction to conductivity $\Delta\sigma/\sigma_0>0$,
we see from Eq. (\ref{S}) that thermodynamical fluctuations
above $T_{\rm c}$ always reduce the overall Seebeck coefficient as
temperature approaches $T_{\rm c}$. So the very sharp maximum in the
Seebeck coefficient of high-$T_c$ materials experimentally observed 
in few papers \cite{few}
seems to be unrelated to fluctuation effects within our simple model
even leaving aside the question about experimental reliability of these
observations. 

Authors are grateful to Collaborative NATO Grant \# 941187 and 
RFFI Grant \# 18878 for support.
One of authors (A. V.) thanks the CARIPLO Foundation
and Landau network for kind hospitality in Pavia.


\begin{thebibliography}{99}
\bibitem{Maki74} K. Maki, J. Low Temp. Phys. {\bf 14}, 419 (1974).
\bibitem{VL90} A. A. Varlamov and D. V. Livanov, Sov. Phys. JETP
{\bf 71}, 325 (1990).
\bibitem{R91} A. V. Rapoport, Sov. Phys. Solid State {\bf 33}, 309 (1991).
\bibitem{K93} D. Kumar, J. Phys. Cond. Matter {\bf 5}, 8227 (1993).
\bibitem{RS94} M. Yu. Reizer and A. V. Sergeev, Phys. Rev. B {\bf 50},
9344 (1994).
\bibitem{ILVY93} L. B. Ioffe, A. I. Larkin, A. A. Varlamov, L. Yu,
Phys. Rev. B {\bf 47}, 8936 (1993).
\bibitem{BDKLV93} A. I. Buzdin, V. V. Dorin, R. A. Klemm, D. V. Livanov and
A. A. Varlamov, Phys. Rev. B {\bf 48}, 12951 (1993).
\bibitem{RV94} M. Randeria and A. A. Varlamov,
Phys. Rev. B {\bf 50}, 10401 (1994).
\bibitem{fv96} F. Federici, A. A. Varlamov, Sov. Phys. JETP {\bf 64}, 497
(1996).
\bibitem{CCRV90} C. Di Castro, C. Castellani, R. Raimondi and A. A. Varlamov
Phys. Rev. B {\bf 42}, 10211 (1990).
\bibitem{few} M. A. Howson {\it et al}, Phys. Rev. B {\bf 41}, 300 (1990);
N. V. Zavaritsky, A. A. Samoilov, and A. A. Yurgens, JETP Letters {\bf 55},
127 (1992).
\end{thebibliography}
\end{document}